\documentclass[aps,twocolumn,secnumarabic,nobalancelastpage,amsmath,amssymb,
nofootinbib]{revtex4-1}

\usepackage{graphicx}
\usepackage{subfigure}
\usepackage{dcolumn}
\usepackage{amsmath}

\numberwithin{equation}{section}

\begin{document}

\title[Clustering Dynamics in Social Amoebae]
{A Monte Carlo Simulation on Clustering Dynamics of Social Amoebae}

\author{Yipeng Yang}
\email{yangyip@missouri.edu}
%\affiliation {Department of
%Mathematics\\ University of Missouri\\ Columbia, MO 65211, USA}

\author{Y. Charles Li}
\email{liyan@missouri.edu} \affiliation {Department of Mathematics\\
University of Missouri\\ Columbia, MO 65211, USA}

%\curraddr{}
\date{\today}

%\thanks{}
%
%\subjclass{Primary 92; Secondary 37}
%\date{}
%
%
%
%\dedicatory{}

\begin{abstract}
A discrete model for computer simulations of the clustering dynamics
of Social Amoebae is presented. This model incorporates the wavelike
propagation of extracellular signaling cAMP, the sporadic firing of
cells at early stage of aggregation, the signal relaying as a
response to stimulus, the inertia and purposeful random walk of the
cell movement. A Monte Carlo simulation is run which shows the
existence of potential equilibriums of mean and variance of
aggregation time. The simulation result of this model could well
reproduce many phenomena observed by actual experiments.
\end{abstract}

\pacs{Valid PACS appear here}

%\keywords{Collective behavior, social amoebae {\it Dictyostelium
%discoideum}, cAMP}

 \maketitle

\section{Introduction}
There are a lot of biological and medical phenomena which have been
proved to be closely related to chemotaxis, such as morphogenesis,
immune response, cancer metastasis, etc. It is crucial to understand
how a cell senses, responds and directs its movement toward a
chemical signal center in chemotaxis. Chemotaxis has been
investigated for many years \cite{Parent99}\cite{van04}, among which
the chemotaxis behavior of eukaryotic microorganism
\emph{Dictyostelium discoideum} provides an ideal example
\cite{Parent08}\cite{Gregor10} because its life cycle incorporates a
number of basic processes that occur throughout developmental
biology.

There are two major paths in the study of Chemotaxis, namely,
experiment \cite{Gregor10}\cite{Song06} and computer simulation
\cite{Segel77}\cite{Mackay78}\cite{Ishii04}\cite{Dallon97}.
Experiment provides the actual behavior of cell movement, and
simulation models can predict much more beyond. A good simulation
model should be able to reproduce as many  features as possible that
are observed in actual experiment. Continuous models are constructed
through fluid dynamics that generally have neat mathematical
representations, see, e.g.
\cite{Segel77}\cite{Sherratt94}\cite{Levine94}\cite{Siegert94}\cite{Hillen09}.
However, many features that are observed through actual experiments
are ignored, such as impulsive releasing of cAMP and wavelike
propagation of signal. In the meanwhile, discrete models have been
proved to be very flexible to incorporate these features.MacKay's
model \cite{Mackay78} provides a basic frame to construct discrete
models, but the threshold of chemotaxis was set below the threshold
of signal relaying, which is not in accordance with recent
experiment \cite{Gregor10}. Dallon and Othmer \cite{Dallon97}
proposed a comprehensive discrete-continuum model that implemented
signal transduction, intracellular and extracellular chemical
exchange, and cell movement along the gradient of cAMP. A thorough
examination of that model was also provided in their paper. Ishii
{\it et al} simulated the binding of cAMP molecules to receptors by
a Monte-Carlo method. Kessler and Levine \cite{Kessler93} modeled
the dynamics of chemotaxis as a set of reaction-diffusion equations
coupled to dynamical biological entities.

In a recent experiment, Gregor {\it et al} showed that the
stochastic pulsing of amoebic cells at early stage of aggregation
plays a critical role in the onset of collective behavior. This
feature is not seen in most of the simulation models which only
focused on the behavior of aggregation but not the onset of
aggregation. In this paper we present a discrete model for the
clustering dynamics of amoebic cell aggregation that takes into
account the sporadic pulsing of cells, wavelike diffusion of
signaling cAMP, signal relaying including signal receiving and
reacting, delay in response time, inertia-like motions and random
oscillation in the cell movement. We showed that the stochastic
pulsing in early stage of aggregation is important for synchronous
firing of the cells, and as a result it enacts the quorum sensing
which triggers and  directs the cell movement.

\section{The Clustering Dynamics of Aggregation}
The life cycle of each amoebic cell starts from a spore. The cell
grows from the spore and divides into more cells as long as there is
food on which they can feed. When food becomes scarce, the cells
begin a different phase of development called aggregation. The
resulting aggregate takes the form of a slug, and the cells
differentiates into a fruiting body after a phase of migration. The
fruiting body consists of a mass of spores surmounting a stalk, and
the dispersal of the spores starts a new life cycle. Interested
readers are referred to Section 5.1 of \cite{Goldbeter96}.

After starvation, small signaling molecules of 3'-5'-cyclic
Adenosine Monophosphate (cAMP) are synthesized and secreted by the
cells into the extracellular space. It is known that the secretion
of cAMP of each cell is not a continuous behavior, but an impulsive
behavior. The release of a pulse of cAMP is about $6\times
10^6-10^7$ molecules \cite{Mackay78}, and this action of a cell is
called a \emph{firing}. This pulse of cAMP travels in the
extracellular space like waves at the speed of 300 $\mu m$/min
\cite{Hofer95}. In the meanwhile, the cells secret Phosphodiesterase
(PDE) that degrades the extracellular cAMP.

At the beginning of this phase, cells fire sporadically once every
15 to 30 min. Over the next 2 hours, the period of firing shortens
to 8 min and thereafter 6 min when the cells begin to aggregate
\cite{Gregor10}. Also when aggregation starts, the firing rate of
the cells is observed to be well synchronized, which is due to
\emph{signal relaying}.

The signal relaying is not detected at earlier stage after
starvation when the concentration of extracellular cAMP is below a
threshold \cite{Gregor10}. At this time the cells move randomly,
fire sporadically, and show a random walk behavior. When the
concentration of the molecules around a cell reaches a threshold
(about $8\times 10^{-9}$ M, \cite{Mackay78}), the cell undergoes a
transition to an oscillatory state in which it can detect the
gradient of cAMP density around it and tends to move to places with
higher concentration of molecules. This behavior is called
\emph{Chemotaxis}, and the overall movement of the population of the
cells behaviors like biased random walk. The cell movement velocity
is found to be about $20-30$ $\mu m$/min \cite{Hofer95}.

Also in the oscillatory state, a cell relays the signal by releasing
a pulse of cAMP in response to a sufficient stimulus. This process
is termed \emph{quorum sensing}. In this state, a cell detects a
wavefront of extracellular cAMP, and after a delay period of about
12-15 s, it releases a pulse of cAMP, and begins the movement step.
After that, the cell enters the refractory period during which it is
insensitive to further stimulus. The refractory period shortens
gradually with age \cite{Robertson75} during which the cell does not
fire. After the refractory period, the cell is ready to respond to
new stimulus if the concentration of molecules is above the
threshold, and if there is no stimulus for a while, the cell fires
autonomously and then enters another refractory period. It is also
found that there is a threshold of the stimulus (wavefront of cAMP)
below which the cell does not respond to the signal \cite{Song06}.

\section{Computer Simulation: Continuous Model {\it vs} Discrete Model}
There have been many models on chemotaxis and clustering behavior of
Dictyostelium discoideum or other bacterial colonies.  The
continuous model certainly plays an important role in this study due
to its neat mathematical representation, see, e.g.
\cite{Segel77}\cite{Sherratt94}\cite{Levine94}\cite{Siegert94}\cite{Hillen09}.
However there are some issues that the continuous models can not
solve easily. (1) The release of cAMP of the cells, either in
response to stimulus or autonomous firing, is impulsive. To simulate
this behavior using continuous models has already been difficult,
see, e.g., Chapter 5 in \cite{Goldbeter96}. (2) It is not easy to
implement the small time delay in the response time of stimulated
firing into the continuous models. (3) The released cAMP propagates
in the media as waves at a certain speed, and most continuous models
lack this feature. (4) In response to stimulus, it is believed that
the cells sense the wavefront of the signal and begin to move. That
is, the cells do not exactly follow the gradient of extracellular
cAMP, it is the wavefront that steers the motion. (5) The
synchronization of firing as a major characteristic of quorum
sensing is not seen in most continuous models. (6) It is well known
that
 there are mainly two ways of movement in chemotaxis: straight swim and
 tumble. As a consequence, the motion of a stimulated cell behaves like a
 purposeful random walk. And if the concentration of molecules is
 below the threshold, the cells just randomly moving. Therefore, to
 implement the randomness into the continuous models could
 dramatically complicate the models.

 Discrete models provide us a lot of flexibility to simulate the
 clustering dynamics of aggregation. MacKay \cite{Mackay78} used a
 discrete model to study the aggregation in Dictyostelium
 discoideum, in which the diffusion of cAMP, signal relaying, time
 delay and random movement were all taken into account. That model is sufficient to
 produce many of the observed slime mould aggregation features, but
 more features can be added, such as the purposeful random walk of cell motion. In
 \cite{Kessler93} the authors modeled this system as a set of reaction-diffusion equations coupled to dynamical biological
 entities, and thus they proposed a framework which is suitable for modeling other biological pattern-forming
 processes. Nishimura and Sasai \cite{Nishimura05} incorporated the inertia into amoebic
 cell locomotion and, through a Monte Carlo simulation, found that
 the averaged position of stimulated cells can be described by a
 second-order differential equation of motion. ``These `inertialike' features suggest the possibility of Newtonian-type motions in chemical distributions of the signaling molecule.''

\section{Simulation Model}
In this section we shall present in detail a discrete model for the
clustering dynamics of cell aggregation. We set up a simulation
environment that is close to the experiment done by Gregor {\it et
al} \cite{Gregor10}. In their work, the authors put about 180 cells
into a 420-$\mu m$-diameter container on hydrophobic agar. In our
simulation, we use a $400\times 400$ grid to represent a $400\times
400$ $\mu m^2$ area. Each time step in the simulation represents 3
seconds, so that in each step, a cell can move 1 $\mu m$ (1 step in
the grid). Since the signal detecting and relaying plays a critical
role in quorum sensing, we first model the wavelike propagation of
the signal.

\subsection{Signal Propagation}\label{sp}
The impulsive release of cAMP includes both the intracellular and
extracellular cAMP, and the extracellular cAMP, which is
comparatively a smaller amount, plays the role of signaling. It is
known that this signal propagates in the media like waves at a speed
of about 300$\mu m$/min. That is 15 steps in the grid in each time
interval.

\begin{figure}[ht]
\centering
\includegraphics[width=9cm,height=2.5cm]{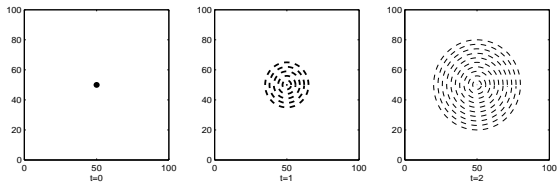}
\caption{Signal Propagation}
\end{figure}
Since the action of PDE and diffusion away from the source cause the
concentration of signal to decrease with distance from the
signalling cell, we propose the following model to represent the
signal strength that a cell receives from the source:
\begin{equation}\label{sig}
s_{i,j}=k_1\frac{r^n}{1+w\cdot d_{i,j}},
\end{equation}
where $i$ is the index of the focused cell, $j$ is the index of the
signalling cell, $k_1$ is a scale factor, $r<1$ is the decaying
factor due to the action of PDE, and $n$ is the time elapses after
the firing, $w$ is a weight factor and $d_{i,j}$ is the Euclidean
distance between cell $i$ and cell $j$. In the simulation we chose
$k_1=1,r=0.95,w=0.038$. When $n>n_0$, the signal from this source is
considered negligible. We chose $n_0=360$ in the simulation
settings.

\subsection{Autonomous Firing}
In Gregor's experiment \cite{Gregor10}, the authors showed that
``the stochastic pulsing of individual cells below the threshold
concentration of extracellular cAMP plays a critial role in the
onset of collective behavior'', and they showed that the firing rate
changes from once every 15-30 minutes to once every 6 minutes. To
simulate this behavior, we first set a base refractory time window
to be 6 minutes (or 120 steps in the simulation). Beyond this
period, an additional time window is added, whose length decreases
with age. The length of this additional time window as a function of
age is given by $twl(n)=ceil(320*(k_2-\arctan(k_3*(n-k_4)/\pi)))$,
where $k_2=0.473,k_3=0.001,k_4=7000$, and $n$ indicates time steps
after starvation. Thus, the length of this additional time window
decreases from 15 minutes at the beginning of simulation to about 0
by the end of the simulation, where $n=20000$ indicating
approximately 16 hours and 36 minutes.

In the first step of simulation, the firing time spot of each cell
is randomly chosen within a 30-min time window. After the first
firing, a cell enters the refractory period whose time length equals
$\max(120,twl(n))$. After this period, the firing time spot is
randomly chosen within the time window $twl(n)$, if the cell is not
stimulated.

\subsection{Signal Relaying}
In each time step (or iteration), a cell accumulates the signal sent
by other cells as described in Section \ref{sp}. This accumulation
is taken as the concentration of extracellular cAMP, and if it is
below a threshold, say $thd_1$, the cell does not respond to it. If
this concentration is above $thd_1$, and the cell is NOT in the
refractory period, the cell responds to this signal by releasing a
pulse of cAMP. This is known as signal relaying and thus the quorum
sensing is formed. It is also known that a cell relays the signal
only if it senses a wavefront of the stimulating signal
\cite{Goldbeter96}. Thanks to the discrete model, this feature can
be easily obtained by taking the difference of current and previous
signal. If this difference is bigger than a threshold \cite{Song06},
$thd_2$, the cell is stimulated and relays the signal. In the
numerical simulation, we set $thd_1=1.1$ and $thd_2=2.5\times
10^{-5}$. A time delay of 5 steps, indicating 15 seconds of time
delay
 in response \cite{Mackay78}, is also incorporated before the release
of cAMP of the stimulated cell.

The simulation code maintains a queue of the positions of all signal
sources. If a cell fires at a certain position, that information is
added into the queue. Associated with each entry of the queue there
is a number indicating the lifetime of that signal. If the lifetime
exceeds $n_0$, that signal is negligible and that entry is
eliminated from the queue.

\subsection{Cell Movement}
When the concentration of extracellular cAMP around a cell is below
$thd_3$ (set at 2.42 in the simulation), the cell moves like random
walk. In this case, a cell has two different ways of movements,
namely the straight swim and tumble. To simulate this feature, we
assume that with half of the chance the cell stays at the same
position in the next  step, and with equal probabilities (1/16) it
moves to any of the neighboring points. In Gregor's experiment
\cite{Gregor10}, the cells were observed to fire synchronously 5
hours after starvation, and over the next 2 hours, the firing rate
shortened to once every 6 minutes and the cells began to move.
Therefore, it is reasonable to make $thd_1<thd_3$, i.e., the
threshold of signal relaying is less than the threshold of
chemotaxis so that synchronous firing precedes chemotaxis. This is
different from the settings in \cite{Mackay78}.

\begin{figure}[ht]
\centering
\includegraphics[width=1.5in,height=1.5in]{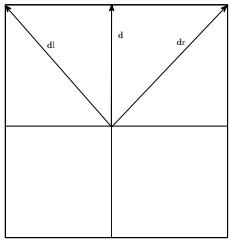}
\caption{Moving directions} \label{M1}
\end{figure}

When the concentration of cAMP is above $thd_3$, the chemotaxis
starts, and a cell moves in a way like purposeful random walk due to
the two different movements. Although the advantage of continuous
model is its capability to incorporate acceleration or deceleration,
it is also possible to implement this feature into the discrete
models by taking into account the inertia \cite{Nishimura05}.

\subsubsection{The Choice of Direction}
We use a vector ${\bf d}$ to indicate the moving direction of a
cell. If a cell is not stimulated, ${\bf d}=0$.  In each iteration,
the cell is assumed to be able to sense the difference of the signal
coming from the neighboring points. Let $s_1\geq s_2$ be the
strengths the two most strongest signal from the neighboring points,
and ${\bf d}_1,{\bf d}_2$ be their respective directions apart from
the cell. $s=s_1{\bf d}_1\cdot s_2{\bf d}_2$ is computed. Notice
that if strong signals coming from opposite directions, $s$ is still
a small number. We take $s$ as the tendency to choose the direction
${\bf d}={\bf d}_1$, and we classify $s$ into `strong', `medium' and
`weak' according to the value of $s$. In each class of $s$, we
assign a number $m_i$ to indicate the number of steps that the cell
tends to move in the direction ${\bf d}$. In the simulation, we set
$m_1=6$, $m_2=4$ and $m_3=2$ associated with the three classes of
$s$, respectively. A 5-step time delay is implemented before the
start of the movement of the cell, and this is the same time when
the stimulated cAMP is released \cite{Mackay78}.

\subsubsection{The Purposeful Random Walk}
After determined the moving direction ${\bf d}$ and the number of
steps to move in that direction, the cell does not necessarily move
straightly in that direction. Due to tumbling, the cell moves along
a direction for a few steps and re-choose the direction. To
incorporate this feature, we assume that in each time step, the cell
moves along the direction ${\bf d}$ with a higher probability, but
also with smaller probabilities to move along the neighboring
directions ${\bf d}_l,{\bf d}_r$ as shown in Fig. \ref{M1}. After
each step of movement, the direction ${\bf d}$ is updated to
indicate the actual direction that the cell is going to follow. When
the anticipated number of steps $m_i$ has reached, and there is no
more stimulation during the movement, set ${\bf d}=0$ to indicate
that the cell loses direction and will follow the random walk as
being not stimulated.

\subsubsection{Acceleration, Deceleration and Steering}
It is very obvious that while a cell is stimulated and performing
the purposeful random walk, there are more signals received during
the movement. The cell is assumed to be able to sense the signal at
each time step. That is, in a certain time step, the cell moves
along the direction ${\bf d}$, and senses new signal $s$ with the
strongest signal coming from direction ${\bf d}_1$. We compare ${\bf
d}$ and ${\bf d}_1$. If the angle between ${\bf d}$ and ${\bf d}_1$
is less than $90^\circ$, we take this new signal $s$ as a
\emph{plus}, and set ${\bf d}={\bf d}_1$ for the next step, also,
according to the strength of $s$ we update the anticipated number of
steps $m_i$. On the contrary, if the angle between ${\bf d}$ and
${\bf d}_1$ is greater than or equal to $90^\circ$, we take this new
stimulus as a \emph{minus}. This simulates the situation that, while
a cell is moving in one direction, it senses a strong signal from
the opposite direction. In this case, we reduce $m_i$ by some steps
according to the strength of $s$. In other words, we reduce the
number of anticipated steps that the cell originally tended to move
in direction ${\bf d}$. Once $m_i$ is zero, ${\bf d}$ is set at
zero, which means the cell is ready to choose a new direction once
there is new stimulus.

\section{Simulation Result}
We first randomly put 180 cells into the chamber to reproduce the
experiment result done by Gregor {\it et al} \cite{Gregor10}. A plus
symbol $+$ in the figure indicates the firing of a stimulated cell,
which lasts for two steps (the release of cAMP is impulsive, but we
draw the star for two steps in the iteration in order to observe
it), and $t$ in the title is the index of time steps, where each
step represents 3 seconds in the actual experiment.

\begin{figure}[ht] \centering
\subfigure{\includegraphics[width=1.6in,height=1.4in]{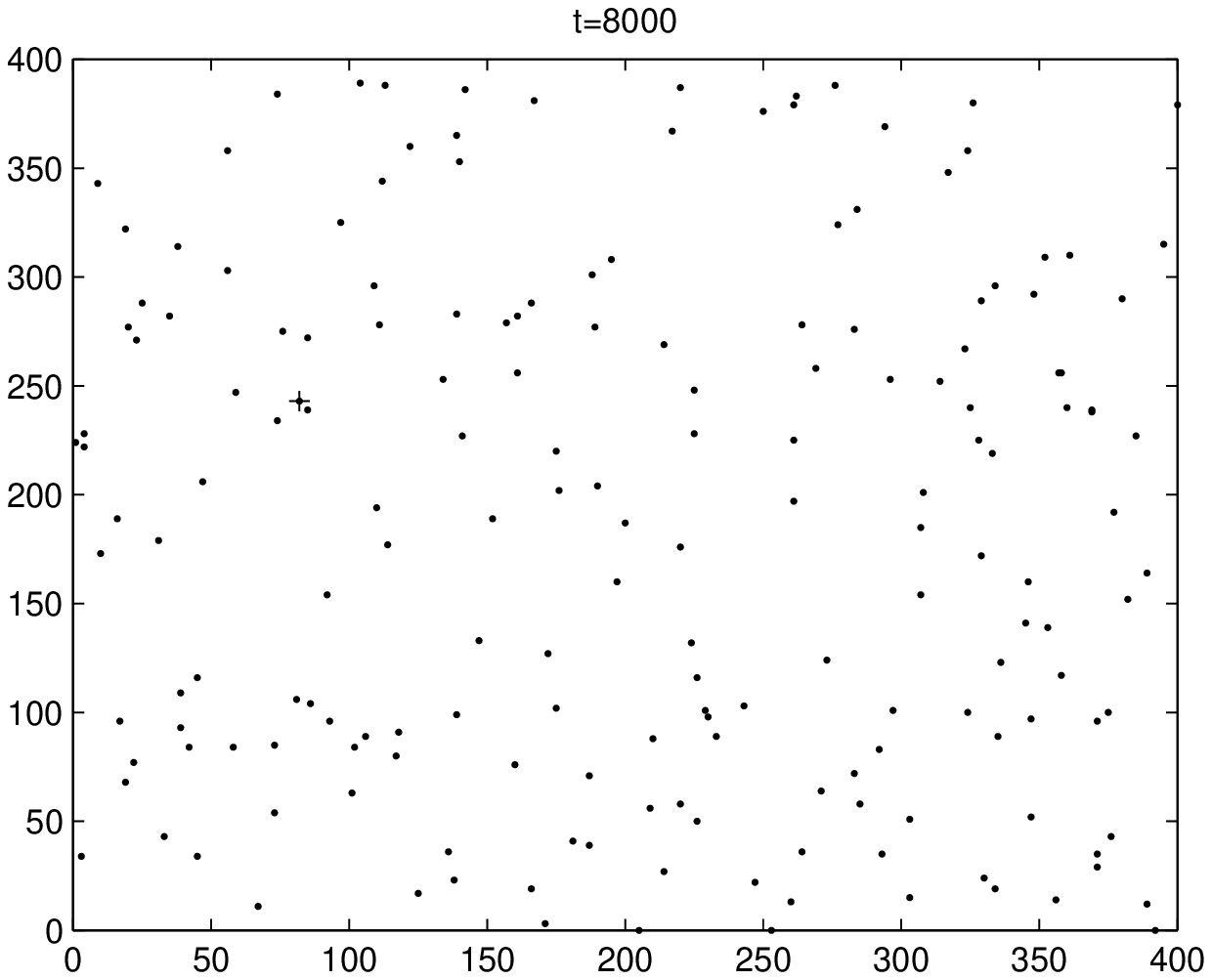}}
\subfigure{\includegraphics[width=1.6in, height=1.4in]{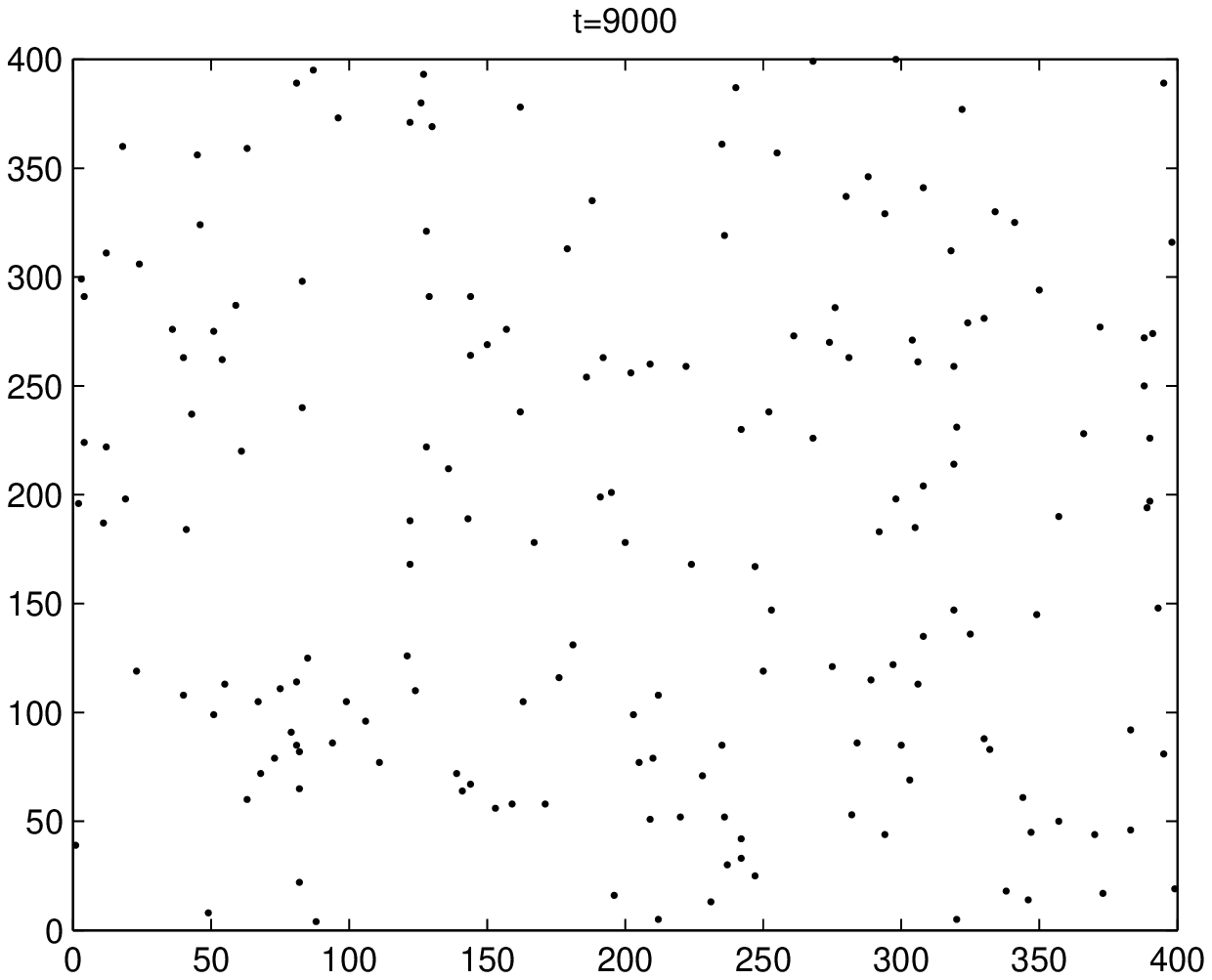}}
\subfigure{\includegraphics[width=1.6in,height=1.4in]{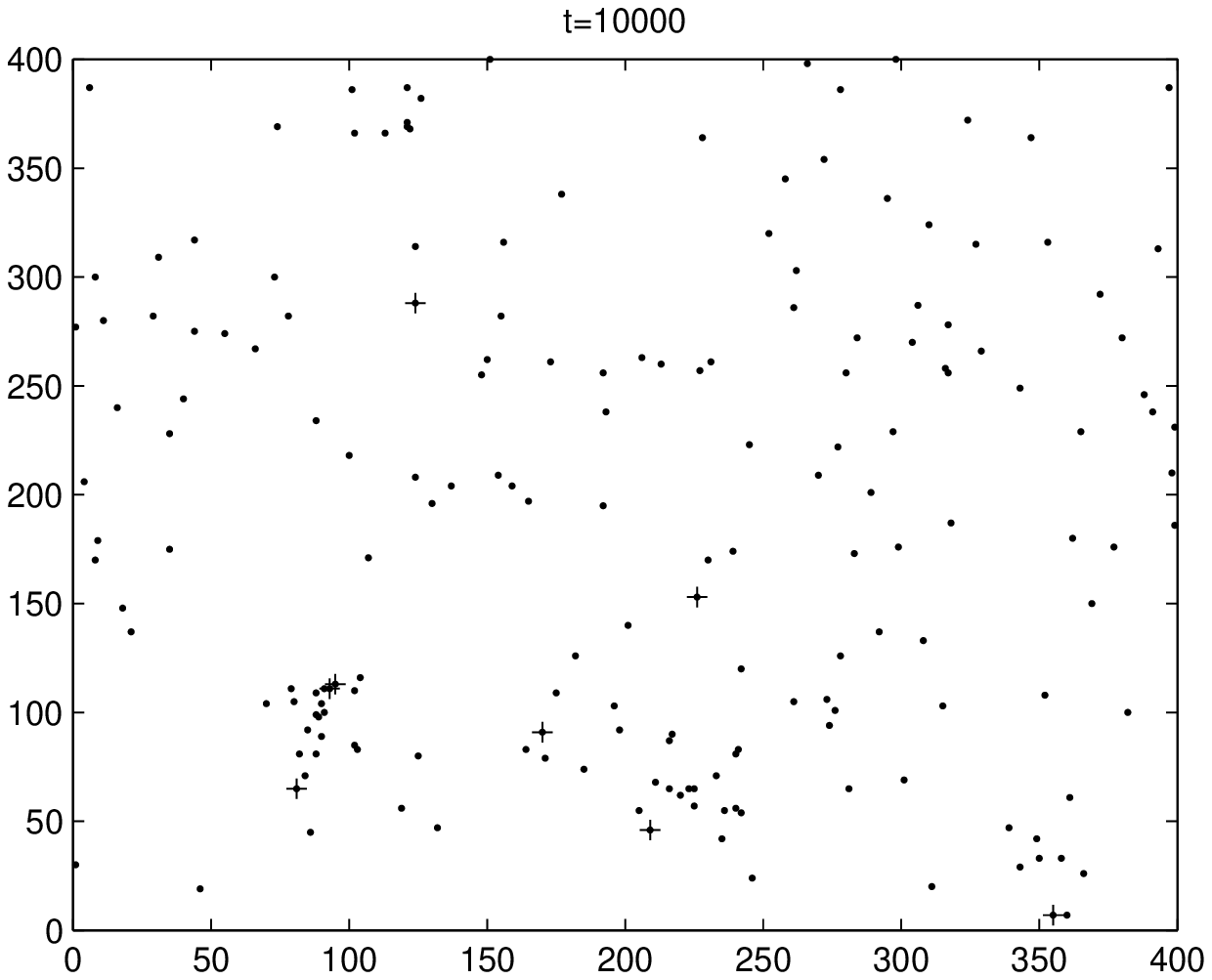}}
\subfigure{\includegraphics[width=1.6in, height=1.4in]{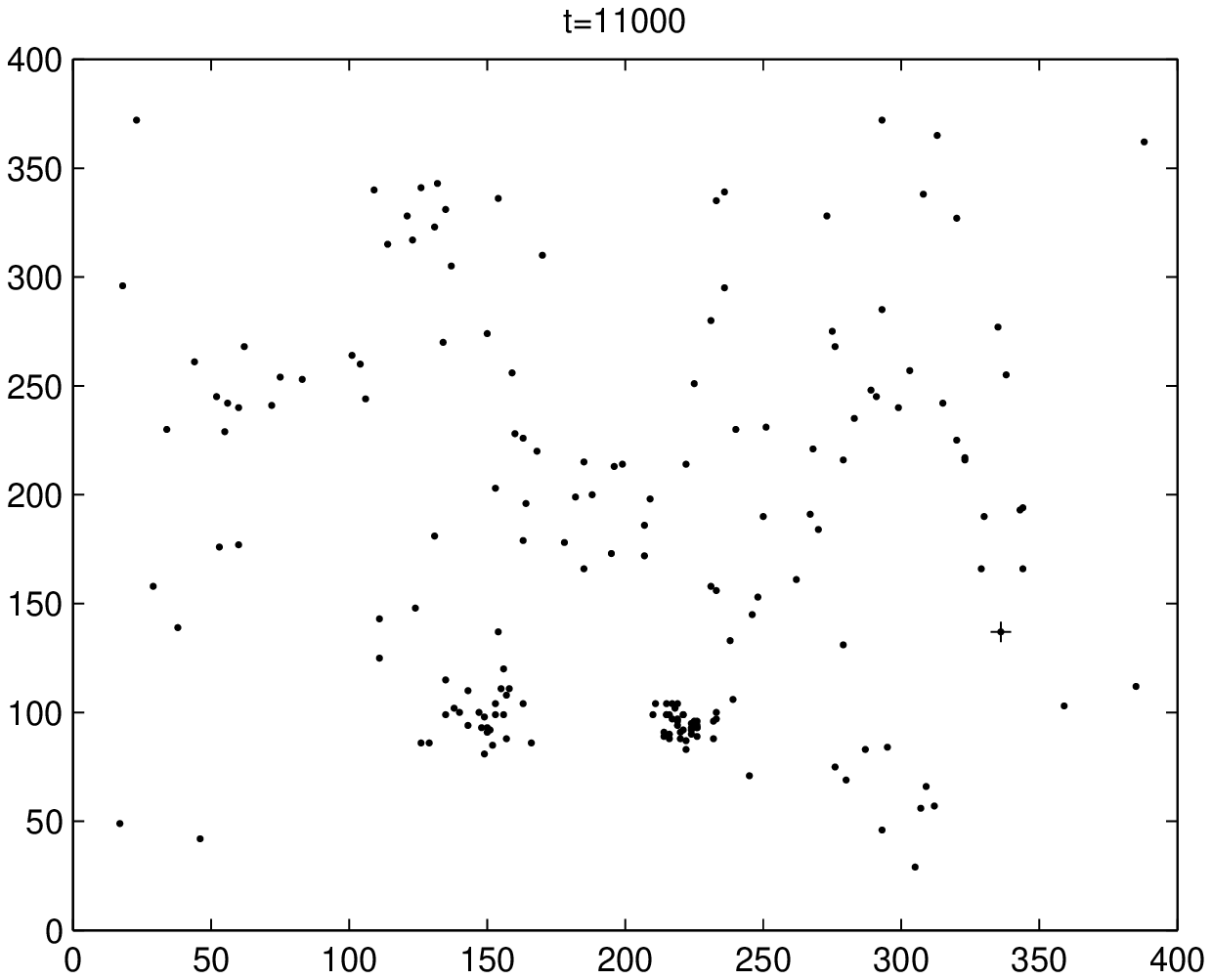}}
\subfigure{\includegraphics[width=1.6in,height=1.4in]{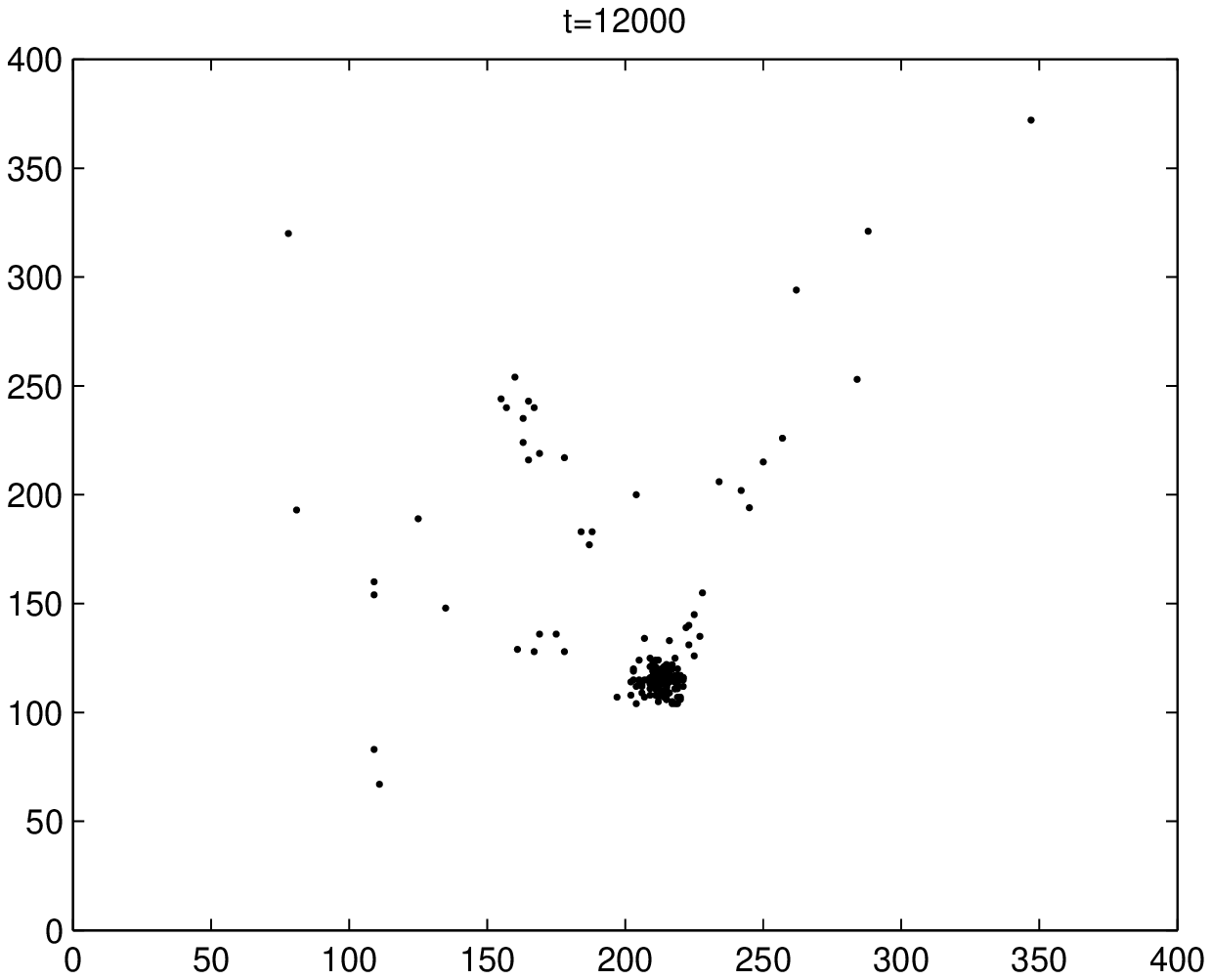}}
\subfigure{\includegraphics[width=1.6in, height=1.4in]{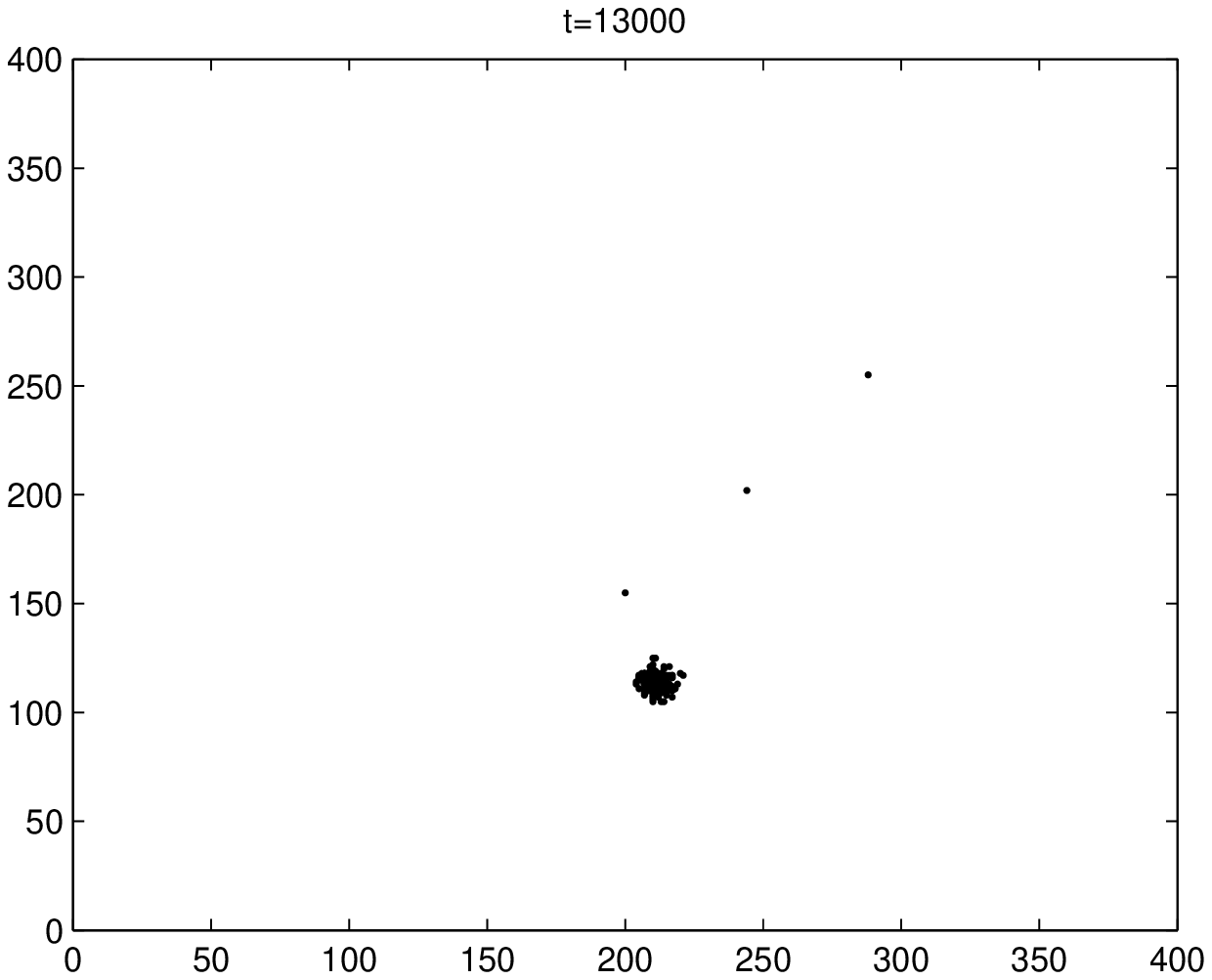}}
\caption{Amoebae Positions, 180 cells} \label{ind1}
\end{figure}
When $t=6000$, synchronous firing is well detected, and when
$t=8000$, the cells began to move. Because of the 15 seconds of
delay in response time, it is not likely to see all the cells
flashing at one particular shot. When $t=11000$ the cells form small
groups which compete, and eventually all the cells aggregate.

\begin{figure}[ht] \centering
\subfigure{\includegraphics[width=1.6in, height=1.4in]{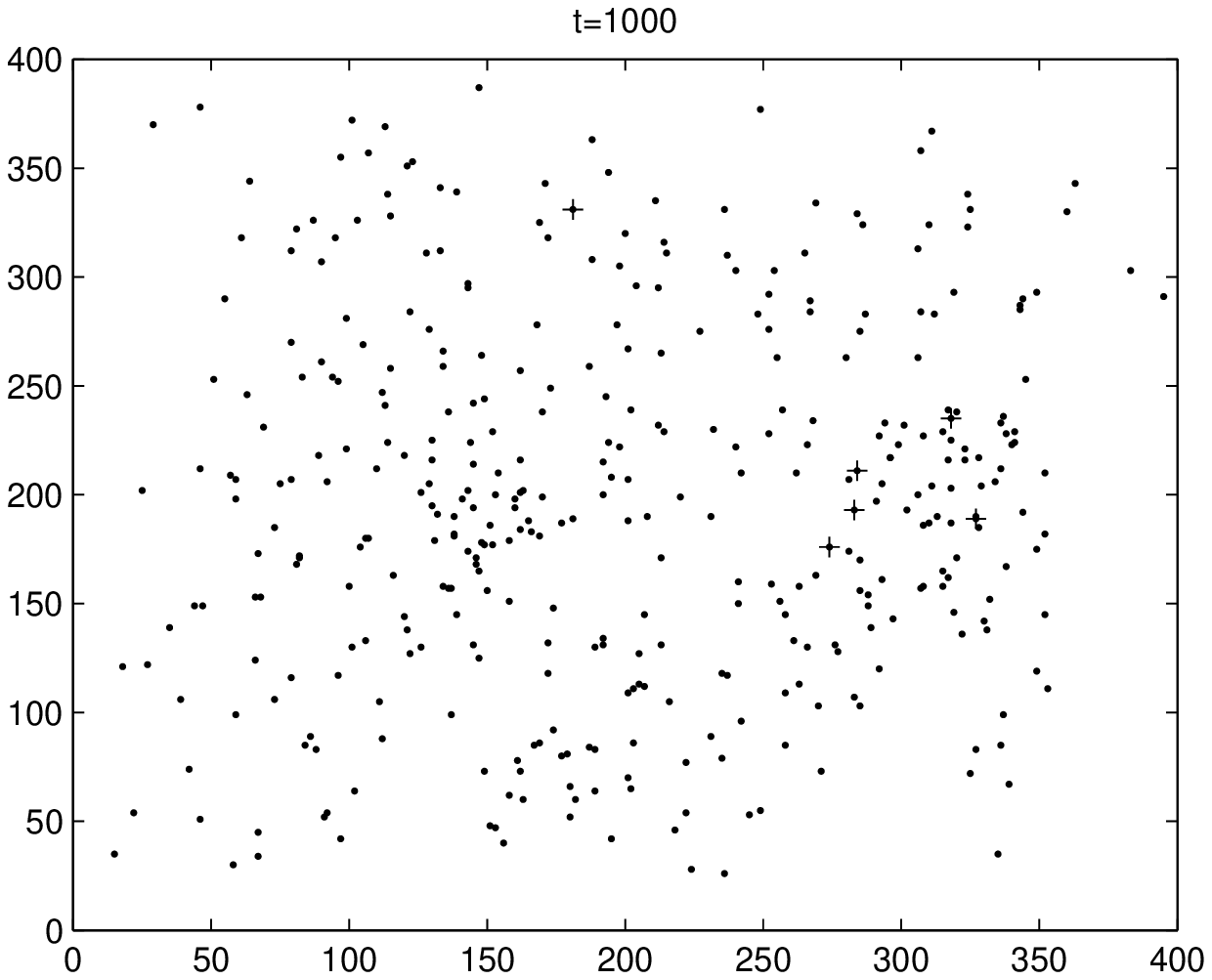}}
\subfigure{\includegraphics[width=1.6in,height=1.4in]{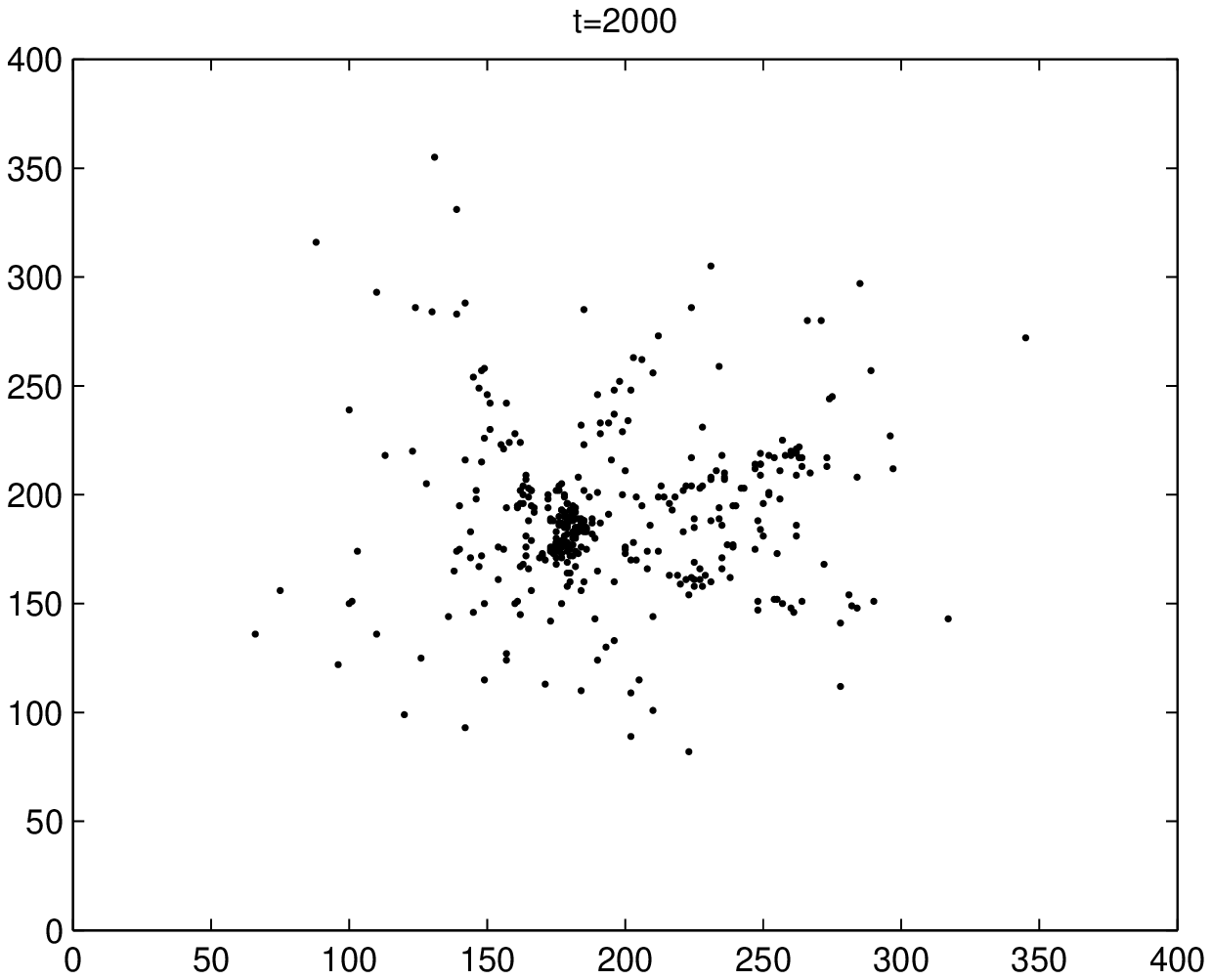}}
\subfigure{\includegraphics[width=1.6in, height=1.4in]{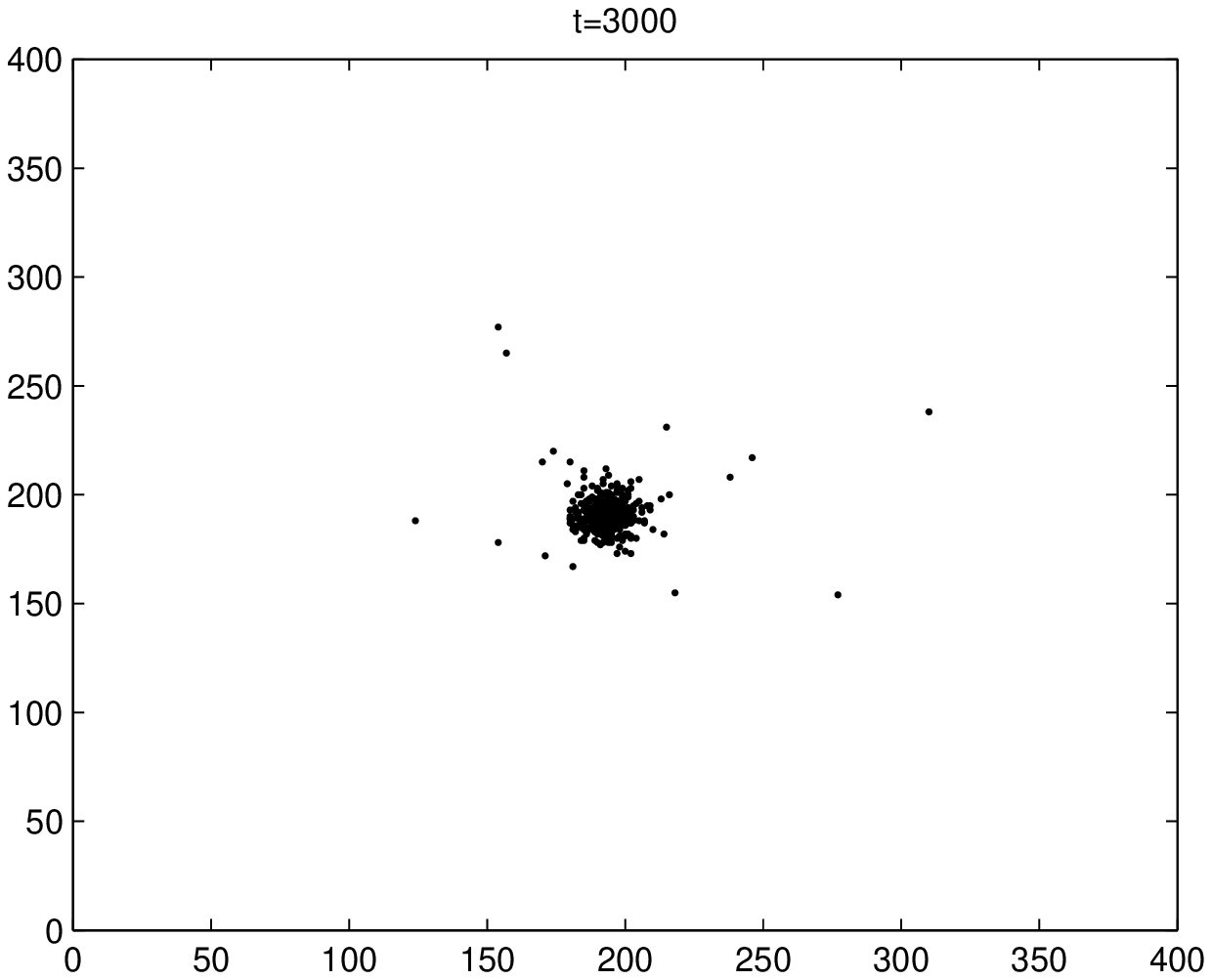}}
\subfigure{\includegraphics[width=1.6in,height=1.4in]{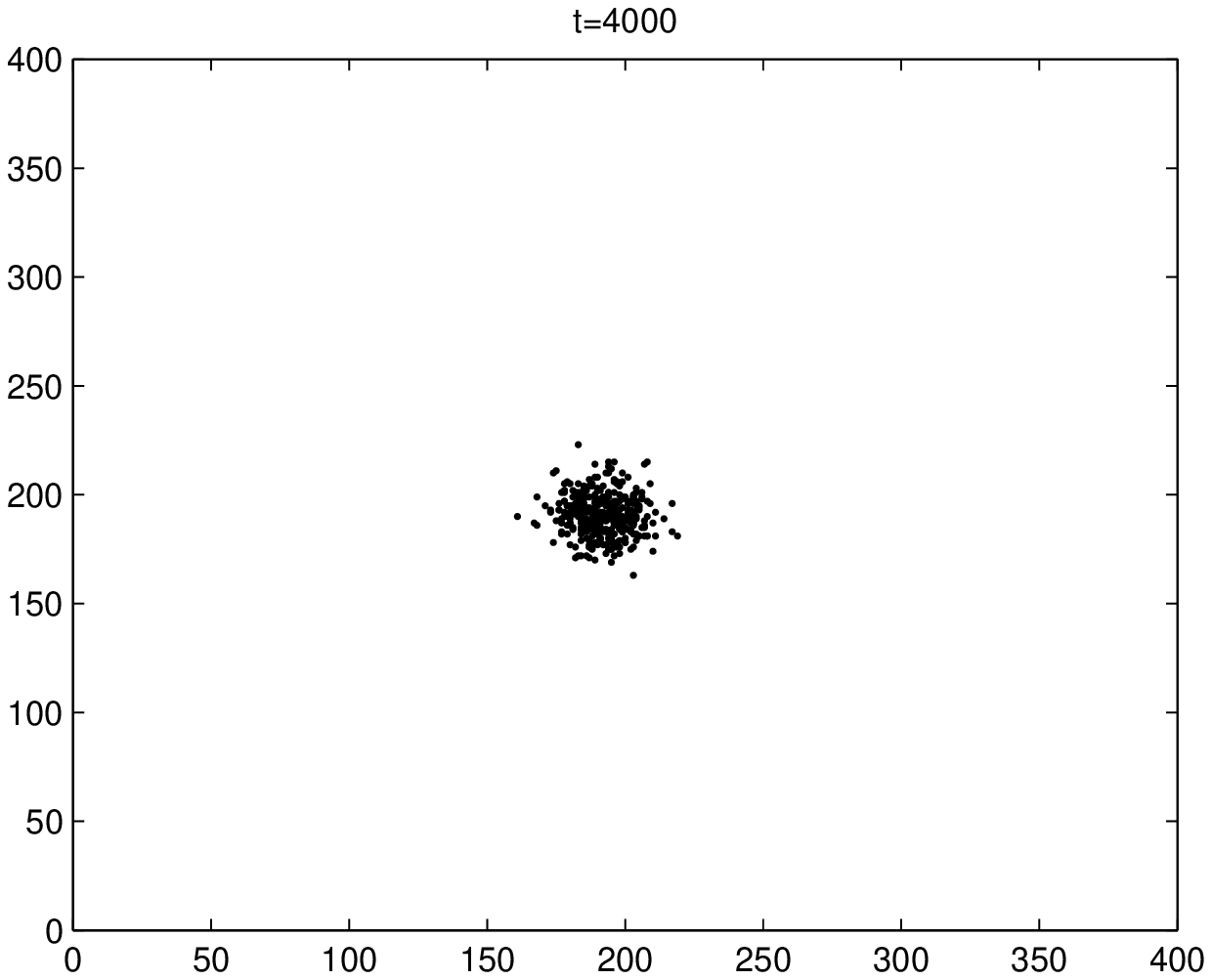}}
\caption{Amoebae Positions, 380 cells}\label{ind2}
\end{figure}

We increase the number of cells to 380 and plot the clustering
dynamics in Figure \ref{ind2}. As expected, more cells increase the
concentration of extracellular cAMP more quickly, and the cells
begin to aggregate at an earlier time.

\begin{figure}[ht] \centering
\subfigure[ 180 cells, 100 trials]{\includegraphics[width=1.6in,
height=1.4in]{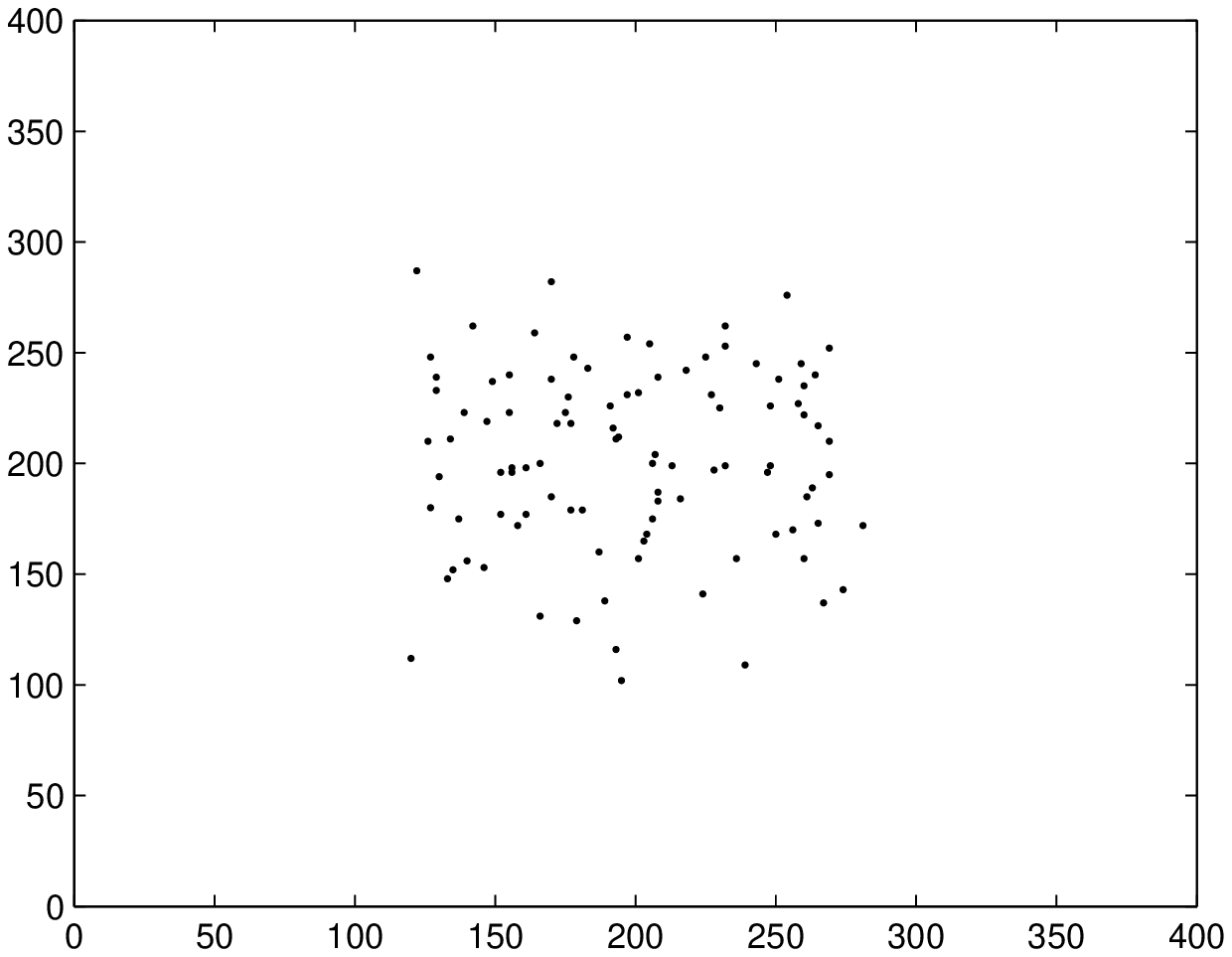}} \subfigure[ 380 cells, 100
trials]{\includegraphics[width=1.6in,height=1.4in]{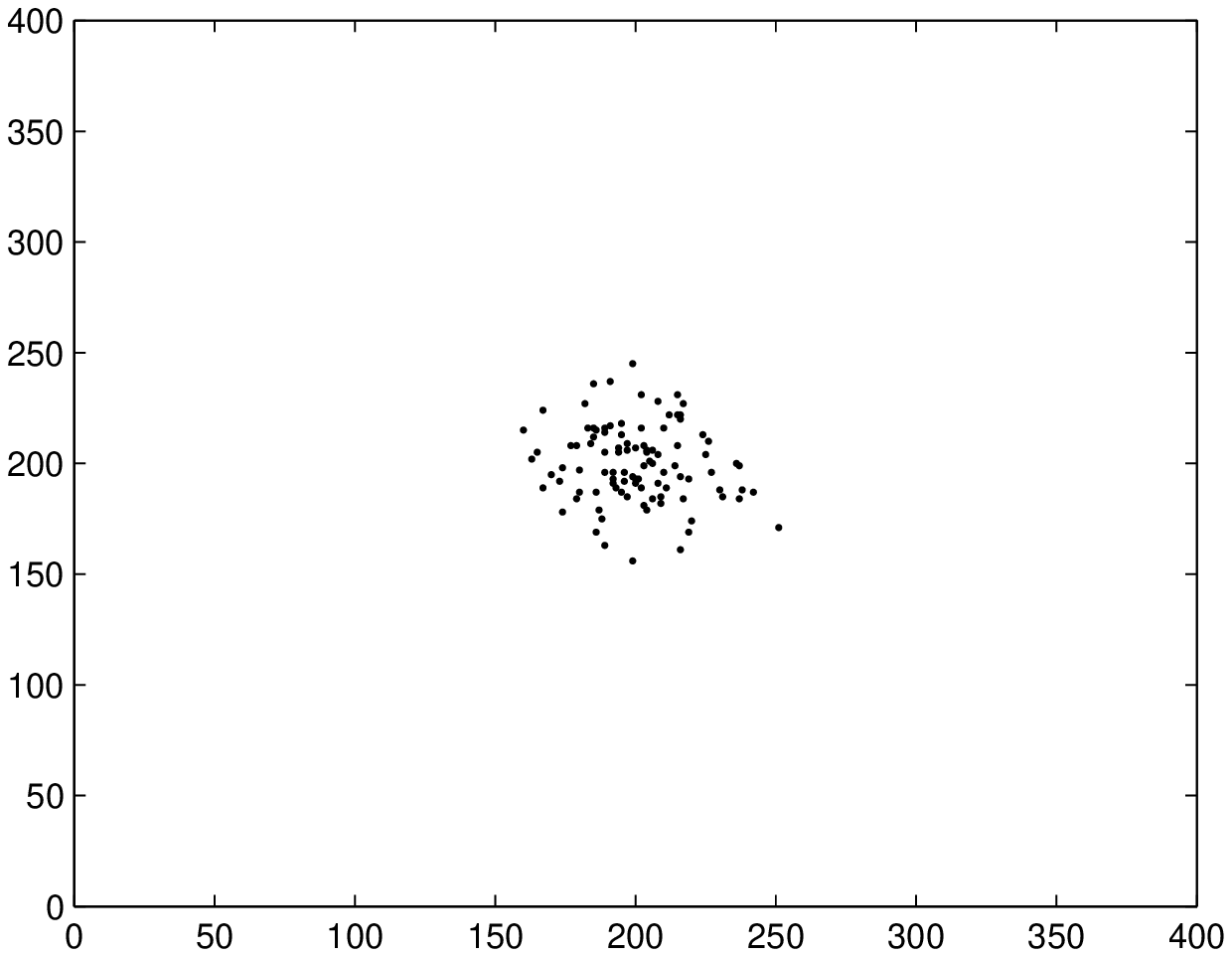}}
\caption{Amoebae Aggregation Centers}\label{centers}
\end{figure}

100 trials are run and the aggregation positions with 180 cells and
380 cells, respectively, are drawn in Figure \ref{centers}.

We can see from Figure \ref{centers} that when there are more cells,
the aggregation center is closer to the center of the chamber, while
less cells make the aggregation center deviated.

\begin{figure}[ht]
\centering
\includegraphics[width=2.2in,height=1.8in]{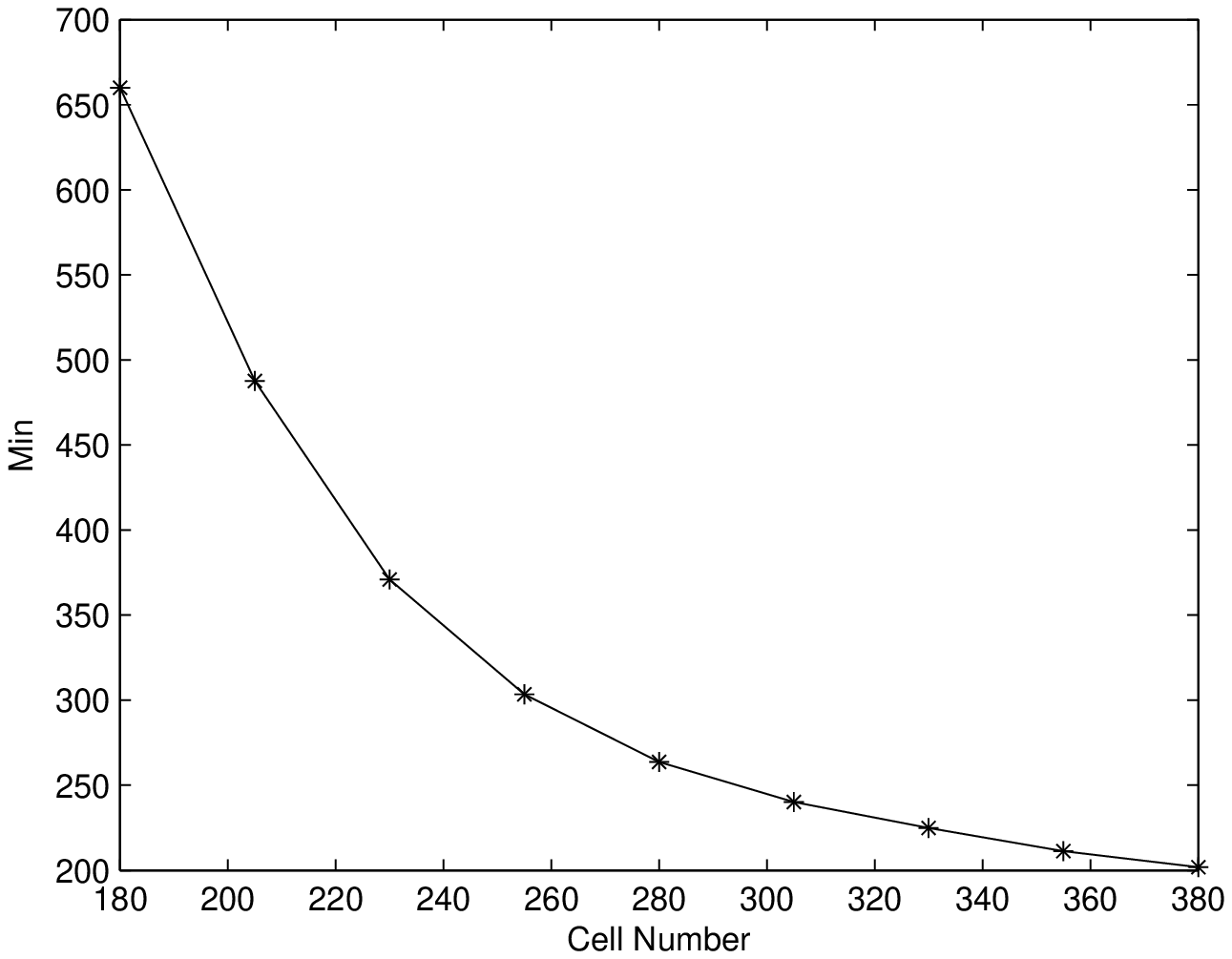}
\caption{Mean Aggregation Time} \label{meantime}
\end{figure}

\begin{figure}[ht]
\centering
\includegraphics[width=2.2in,height=1.8in]{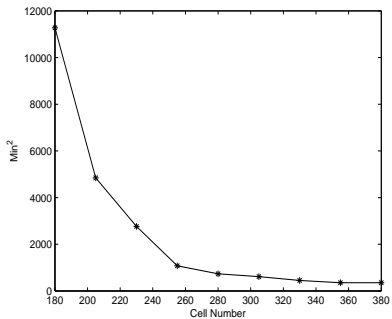}
\caption{Variance of Aggregation Time} \label{vartime}
\end{figure}

\begin{figure}[ht]
\centering
\includegraphics[width=2.2in,height=1.8in]{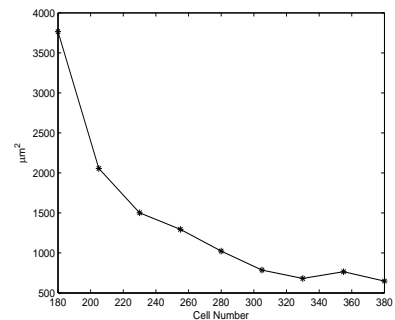}
\caption{Variance of Aggregation Location} \label{varloc}
\end{figure}

The number of cells is increased from 180 to 380 by 25 each time.
For each setting, 100 trials are run and the average aggregation
time, variance of aggregation time and variance of aggregation
location are plotted in Figures \ref{meantime}, \ref{vartime} and
\ref{varloc}, respectively.

As expected, more cells shortened the aggregation time. However,
after the point 280, the curves in Figures \ref{meantime},
\ref{vartime} and \ref{varloc} all flatten out. Since synchronous
firing plays the key role in quorum sensing, the procedure from
sporadic firing to synchronous firing is important for the
clustering dynamics of social Amoebae. Even though more cells
increase the extracellular cAMP more quickly, they do not
necessarily aggregate sooner if the firing are not well paced.
Gregor {\it et al} \cite{Gregor10} showed that the sporadic firing
below the threshold concentration of cAMP plays a critical role in
the onset of the collective behavior of social Amoebae, and we
believe that the role which sporadic firing plays is to make the
firing well paced. Therefore, the way the signal propagates in the
media is critical in the clustering dynamics, and this feature is
often missed in most continuous models.

To further study the effect of signal propagation, we set the number
of cells at 380 and set the parameter $w=0.08$ in (\ref{sig}). This
mimics the situation where the signal decays faster with distance
from the source.

\begin{figure}[ht] \centering
\subfigure{\includegraphics[width=1.6in,height=1.4in]{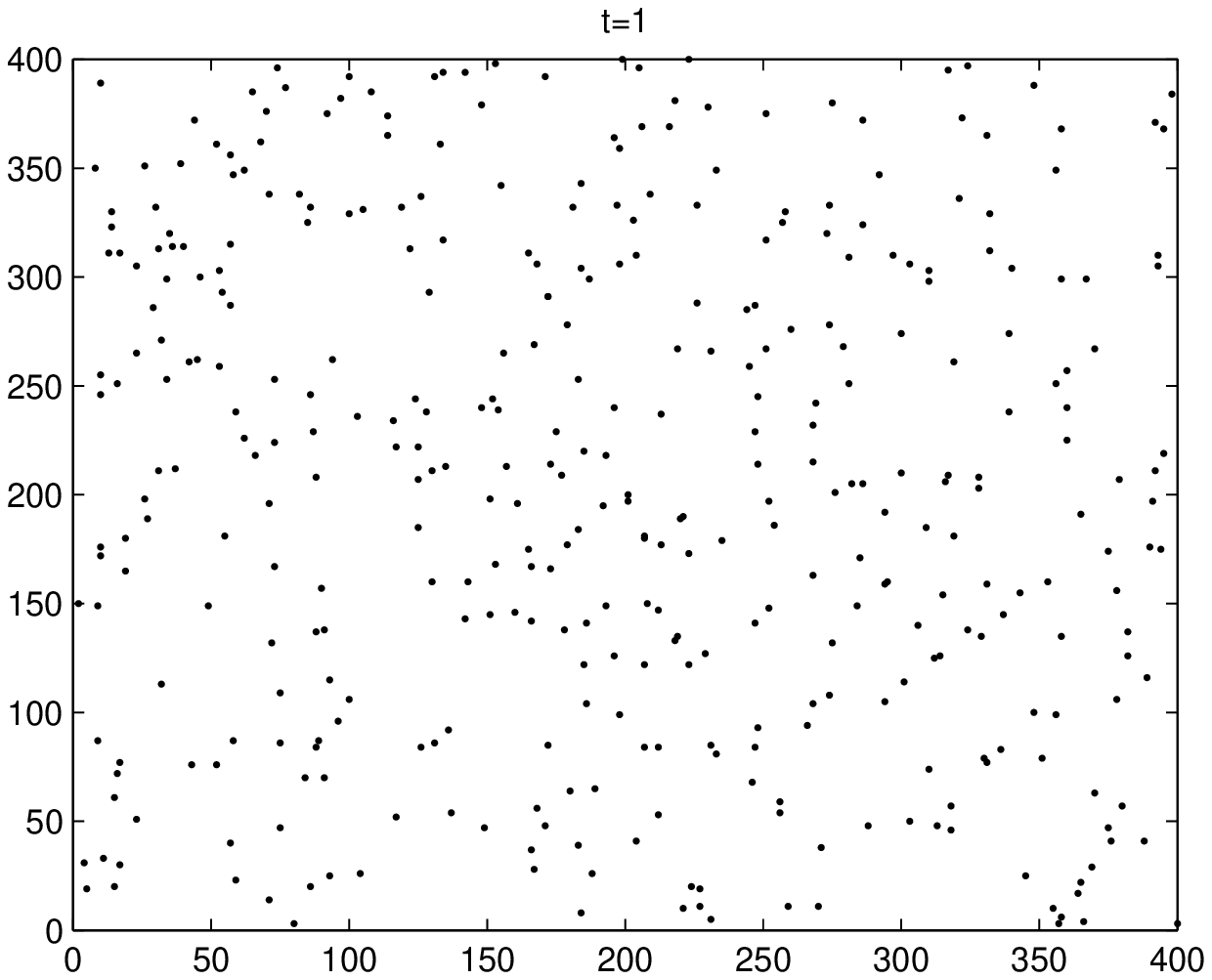}}
\subfigure{\includegraphics[width=1.6in,
height=1.4in]{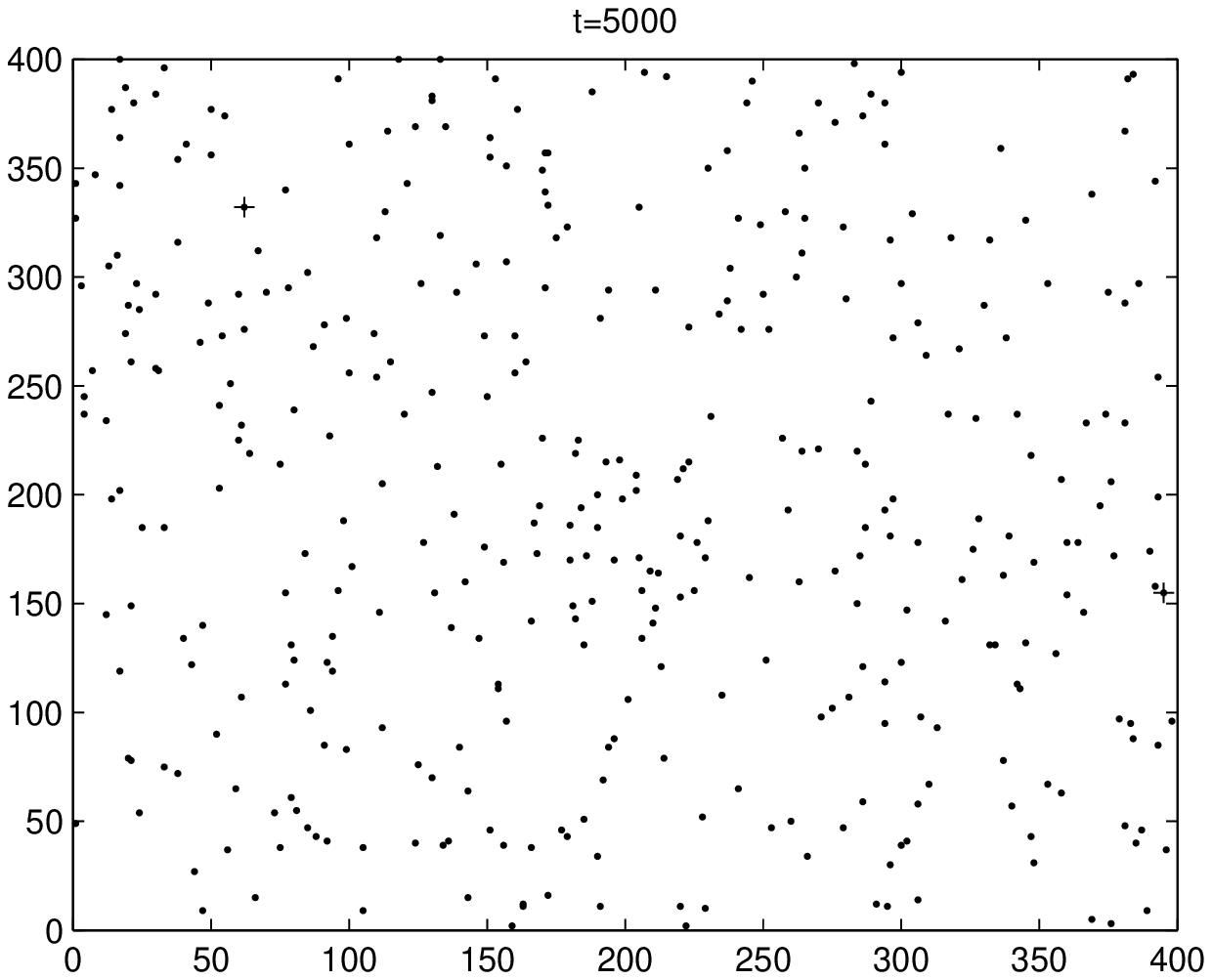}}
\subfigure{\includegraphics[width=1.6in,height=1.4in]{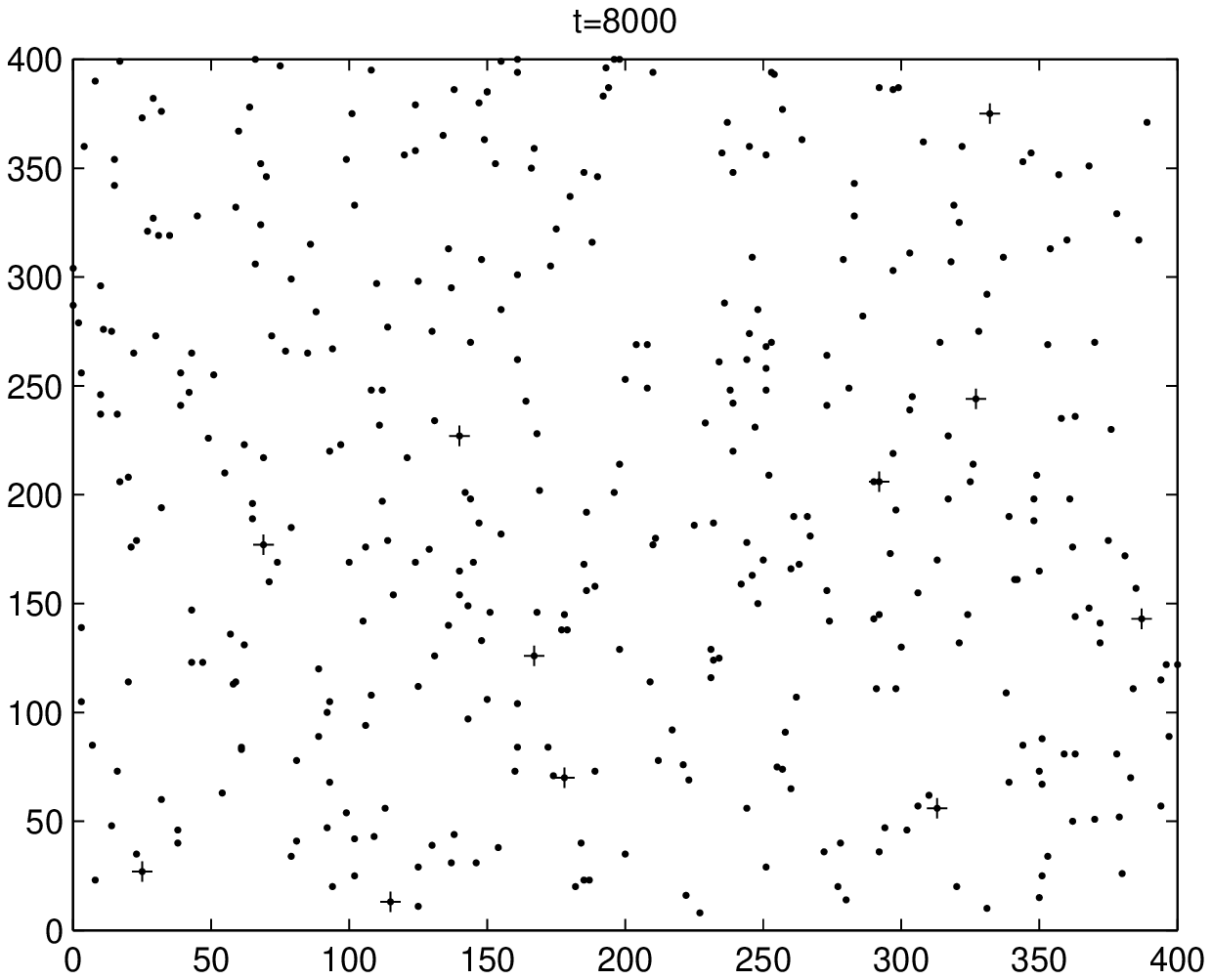}}
\subfigure{\includegraphics[width=1.6in,
height=1.4in]{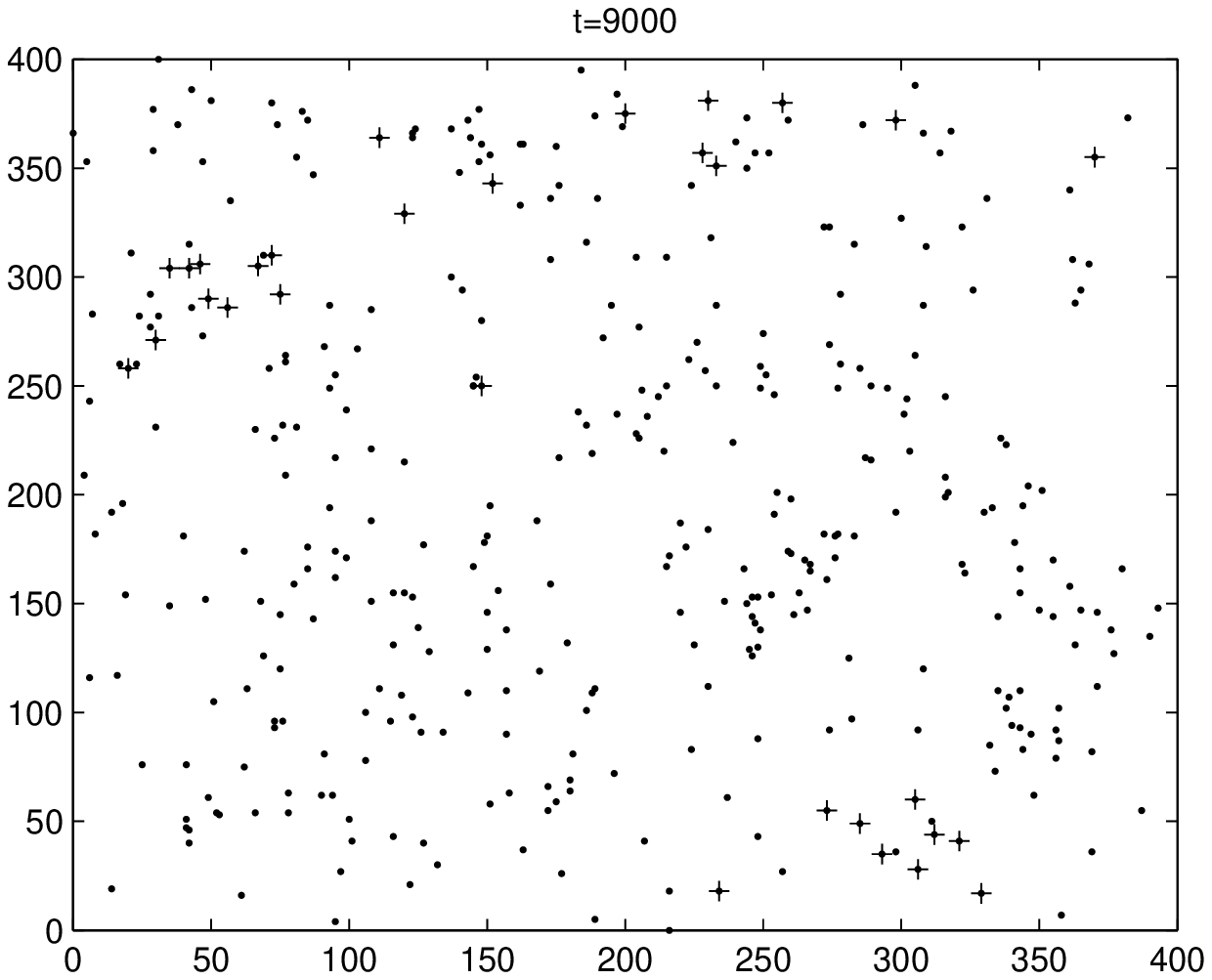}}
\subfigure{\includegraphics[width=1.6in,height=1.4in]{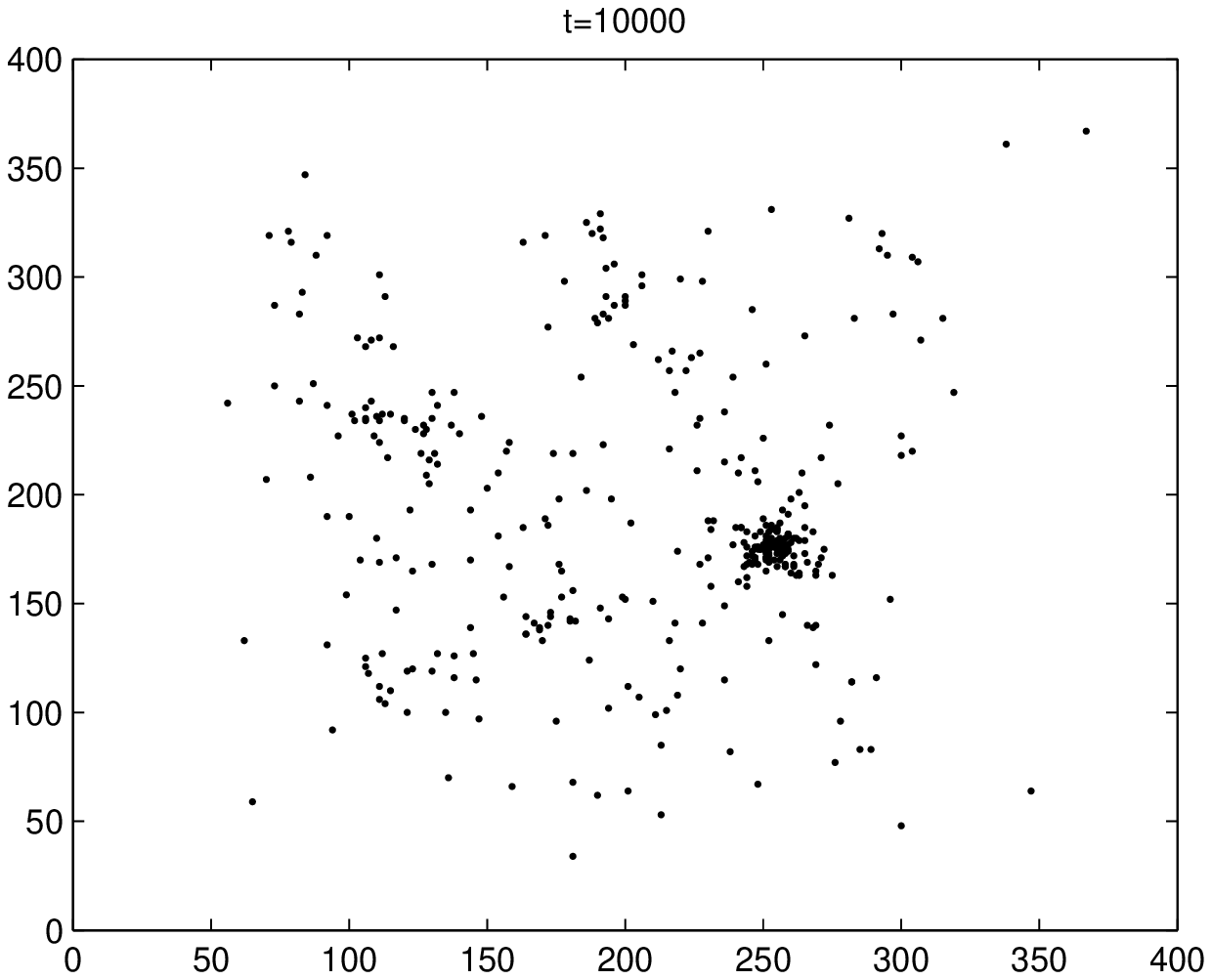}}
\subfigure{\includegraphics[width=1.6in,
height=1.4in]{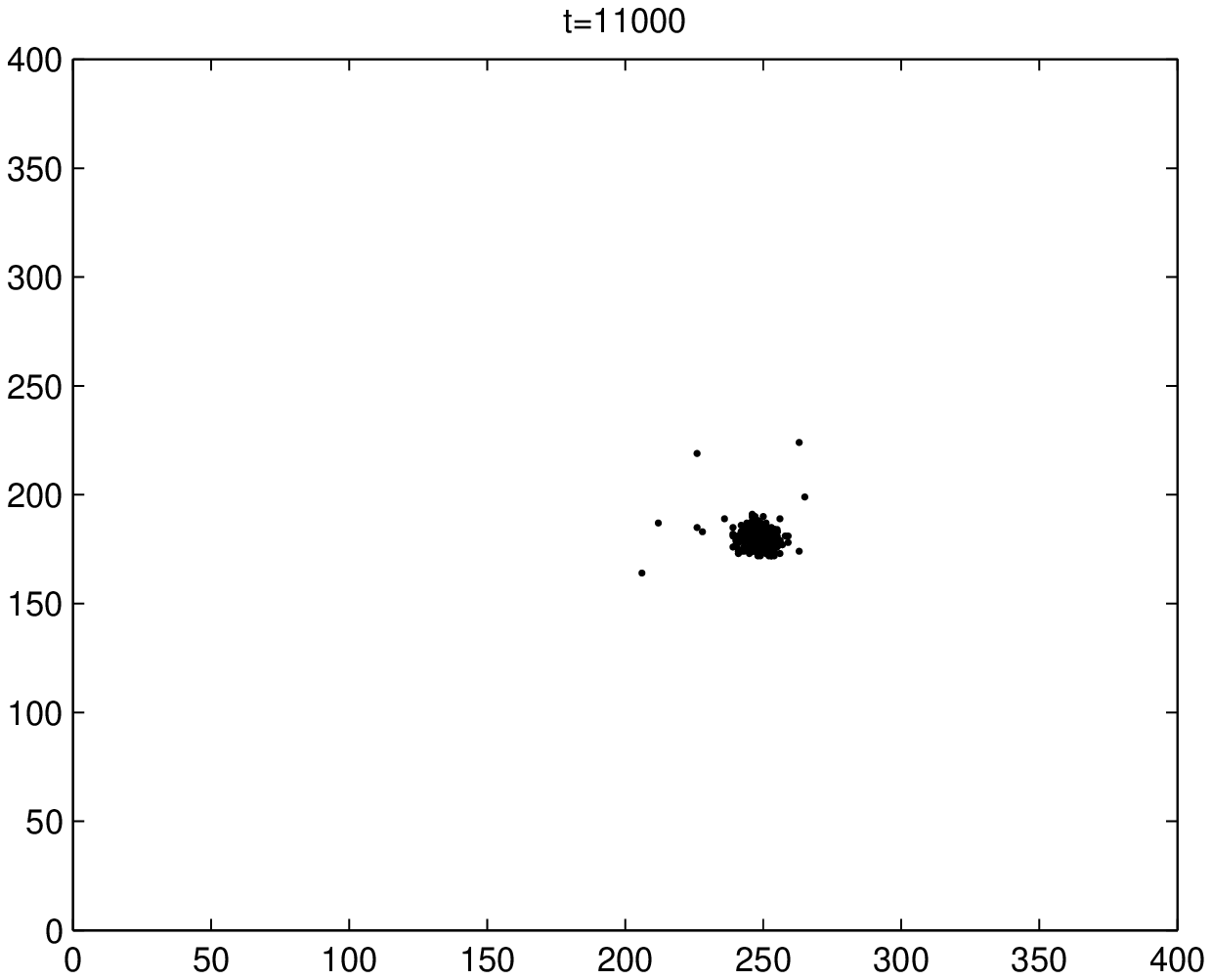}} \caption{Amoebae Positions, 380 cells,
$w=0.08$} \label{ind380n}
\end{figure}

As can be seen from Figure \ref{ind380n}, the cells tend to form
small groups that compete and the firing are not well paced, and as
a consequence, the cells take longer time to aggregate.

\section{Concluding Remarks}
In this paper, we present a discrete model for computer simulations
of the clustering dynamics of Social Amoebae. This model
incorporates the wavelike propagation of extracellular signaling
cAMP, the sporadic firing of cells at early stage of aggregation,
the signal relaying as a response to stimulus, the inertia and
purposeful random walk of the cell movement. The simulation result
of this model could well reproduce the phenomenon observed by actual
experiments. We found that synchronous firing plays a critical role
in the clustering dynamics of social Amoebae. The sporadic firing at
early stage after starvation is necessary to obtain the synchronous
firing preceding the aggregation. A Monte Carlo simulation is run
which shows the existence of potential equilibriums of mean and
variance of aggregation time.

As a future research topic, it might be interesting and necessary to
take into account the volume of each cell, and the change of shape
as a cell moves \cite{Nishimura05}. In that case, we anticipate that
the cells are easier and faster to aggregate.

\end{document}